# EVOLUTIONARY NEURAL GAS (ENG): A MODEL OF SELF ORGANIZING NETWORK FROM INPUT CATEGORIZATION.

Ignazio Licata (a) [†] , Luigi Lella (b)
(a) Ixtucyber for Complex Systems, Marsala, TP and Institute for Scientific Methodology, Palermo, Italy;
(b) A.R.C.H.I. - Advanced Research Center for Health Informatics, Ancona, Italy

ABSTRACT
   Despite their claimed biological plausibility, most self organizing networks have strict topological constraints and consequently they cannot take into account a wide range of external stimuli. Furthermore their evolution is conditioned by deterministic laws which often are not correlated with the structural parameters and the global status of the network, as it should happen in a real biological system. In nature the environmental inputs are noise affected and "fuzzy". Which thing sets the problem to investigate the possibility of emergent behaviour in a not strictly constrained net and subjected to different inputs.
   It is here presented a new model of Evolutionary Neural Gas (ENG) with any topological constraints, trained by probabilistic laws depending on the local distortion errors and the network dimension. The network is considered as a population of nodes that coexist in an ecosystem sharing local and global resources.
   Those particular features allow the network to quickly adapt to the environment, according to its dimensions. The ENG model analysis shows that the net evolves as a scale-free graph, and justifies in a deeply physical sense- the term "gas" here used.

1. INTRODUCTION

   Self organizing networks are systems widely used in categorization tasks. A network can be seen as a set A={c1, c2,… ,cn} of units with associated reference vectors $w_c \in R^n$ where $R^n$ is the same space where inputs are defined. Each unit (or node) can establish connections with the other ones, the units belonging to the same clusters are subjected to similar modification affecting their reference vectors.
   Self organizing networks can automatically adapt to input distributions without supervision by means of training algorithms that are simple sequences of deterministic rules. *Competitive hebbian learning* and *neural gas* are the most important strategies used for their training.
Neural gas algorithm (Martinetz T.M. and Schulten K.J., 1991) sorts the network units according to the distance of their reference vector to each input. Then the reference vectors are adapted so that the ones related to the first nodes in the rank order are moved more close than the others to the considered input.
   Competitive hebbian learning (Martinetz and Schulten, 1991; Martinetz, 1993) consists in augmenting the weight of the link connecting the two units whose reference vectors are closest to the considered input (the two most activated units).
Both strategies are examples of deterministic rules. As we know there are other rules that constrain the topology of the network which has a fixed dimensionality. That's the case of Self Organizing Maps (Kohonen, 1982) and Growing Cell Structures (Fritzke, 1994).

---

[†] Corresponding author: Ignazio.licata@ejtp.info

In other cases the network structures haven't topological constraints, they take a well ordered distribution by exactly adapting to the manifold inputs. For example TRN (Martinetz and Schulten, 1994) and GNG are networks whose final structure is similar to a *Delaunay Triangulation* (Delaunay, 1934).We have tried to define a new self organizing network that is trained by probabilistic rules avoiding any topological constraints.

According to Jefferson (1995) life and evolution are structured at least into four fundamental levels: *molecular*, *cellular*, *organism* and *population*. We propose a population level based on evolutionary algorithm where the network is seen as a population of units whose interactions are conditioned by the availability of resources in their ecosystem. The evolution of the population is driven by a selective process that favours the fittest units. This approach has a biological plausibility. As stated by recent theories (Edelman, 1987) human brain evolution is subjected to similar selective pressures.

Obviously we are not interested in recreating the same structure as the human brain. Our work aims at finding innovative and effective solutions to the categorization problem adopting natural system strategies. So our system falls within the *Artificial Life* field (Langton, 1989).

Our model is a complex system that shows emergent features. In particular its structure evolves as a scale free graph. In the training phase there arise clusters of units with a limited number of nodes that establish a great number of links with the others.

Scale free graphs are a particular structure that is really common in natural systems. Human knowledge, for instance, seems to be structured as a scale free graph (Steyvers, Tenenbaum 2001). If we represent words and concepts as nodes, we'll find that some of these are more connected than the others.

Scale free graphs have three main features.The *small world* structure. It means there is a relatively short path between any couple of nodes (Watts, Strogatz, 1998).The inherent tendency to *cluster* that is quantified by a coefficient introduced by Watts and Strogatz. Given a node i of $k_i$ degree i.e. having $k_i$ edges which connect it to $k_i$ other nodes, if those make a cluster, they can establish $k_i(k_i-1)/2$ edges at best. The ratio between the actual number of edges and the maximum number gives the clustering coefficient of node i. The clustering coefficient of the whole network is the average of all the individual clustering coefficients.

Scale free graphs are also characterized by a particular degree distribution that has a *power-law tail* $P(k) \sim k^{-n}$. That's why such networks are called "scale free" (Albert, Barabasi, 2000).

The three previous features are quantified by three parameters: the *average path length* between any couple of nodes, the *clustering coefficient* and the *exponent* of the power law tail. We'll show that the values of these parameters in our model seem to confirm its scale free nature.

2. AN OUTLINE ON SELF-ORGANIZATION AND EVOLUTIONARY SYSTEMS

Natural selection mechanism has been successfully used for a lot of industrial applications spanning from projecting to real-time control and neural networks training.

It was in the 60s that Genetic Algorithms based on the Evolution Theory's three main mechanisms - *reproduction*, *mutation* and *fitness* – were first used in dealing with optimization problems. Although the solution is reached by a population of individuals, systems based on this approach are not considered self organizing because their dynamics depend on the external constraint of the fitness function.

In the 80s a new approach to the study of living systems which mixed together self organization and evolutionary systems came out (Rocha, 1997). Its success was due to the studies on the way how biological systems work (metabolism, adaptability, autonomy, self repairing, growth, evolution etc.). The hybrid systems make us possible to get a better simulation both of the evolutionary optimization processes and the internal structure modification to reach a greater biological plausibility in the fitness

*Neuroevolutionary* systems are an example of this approach. In classic neuroevolutionary models the network parameters are genetically set, whereas the connection weights are modified according to a training strategy. This solution follows the classic vision of cerebral development where genes control the formation of synaptic connections while their reinforcement depends on neural activity.

More recent neuroevolutionary systems are characterized by different forms of self organizing processes which are *cooperative coevolution* (Paredis, 1995; Smith, Forrest and Perelson, 1993) and *synaptic Darwinism* (Edelman, 1987).

Cooperative co evolutionary systems offer a promising alternative to classic evolutionary algorithms when we face complex dynamical problems. The main difference with respect to classic EA is the fact that each individual represents only a partial solution of the problem. Complete solutions are obtained by grouping several individuals. The goal of each individual is to optimize only a part of the solution, cooperating with other individuals that optimize other parts of the solution. It is so avoided the premature convergence towards a single group of individuals. An example of such approach is given by the Symbiotic Adaptive Neuroevolution System (Moriarty and Miikkulainen, 1998) that operates on populations of neural networks.

While in most neuroevolutionary systems each individual represents a complete neural network, in SANE each individual represents a hidden unit of a two-layered network. Units are continuously combined and the resulting networks are evaluated on the basis of the performances shown in a given task. The global effect is equal to schemas promoting in standard EAs. In fact during the evolution of the population the neural schemas having the highest fitness values are favoured and the possible mutations in the copies of these schemas don't affect the other copies in the population.

Other recent strategies focus on the evolution of connection schemas in the network. In the human brain the number of synapses established by a single neuron is always much lower than the overall number of neurons. That gives the network a sparsely connected aspect. In the last years several models have been proposed to emulate the mechanism involved in the selection of links without referring to the physical and chemical properties of neurons.

The Chialvo and Bak model (Chialvo and Bak, 1999) is based on two simple and biological inspired principles. First, the neural activity is kept low selecting the activated units by a *winner takes all* strategy. Second, the external environment gives a *negative feedback* that inhibits active synapses if the network behaviour is not satisfying. With these simple rules the model operates in a highly adaptive state and in critical conditions (*extreme dynamics*). The fundamental difference of this strategy based on the *synaptic inhibition* with respect to the classic one based on synaptic reinforcement is that the reinforcement-based learning is a continuative process by definition, while the inhibition-based learning stops when the training goal is achieved. The synaptic inhibition is also biologically plausible. According to Young (Young, 1964; Young, 1966) learning is the result of the elimination of synaptic connections (*closing of unneeded channels*). Dawkins (Dawkins R., 1971) stressed that pattern learning is achieved by synaptic inhibition. As stated by the *neural groups' selection theory* developed by Edelman (Edelman, 1978; Edelman, 1987), brain development is characterized by generating a structural and dynamical variability within and between populations of neurons, by the interaction of the neural circuit with the environment and by the differential attenuation or amplification of synaptic connections. Research in neurobiology seems to confirm the validity of the negative feedback model and the fact that neural development follows the process of Darwinian evolution.

The Chialvo and Bak model is a simple two-layered network. After the training each input pattern is associated with a single output unit leading to the formation of an associative map. When an input pattern is presented the most activated input unit i is selected. Then the neuron j from the hidden layer that establishes the most robust connection with i is selected. Finally the output neuron k that is the most strongly connected with j is selected. If k is not the desired output the two links connecting i with j and j with k are inhibited by a coefficient d that is the only parameter of the model. The iterative application of these rules leads to a rapid convergence towards any input-output mapping. This selective process followed by an inhibitory one is the essence of the natural selection in the evolutionary context. The fittest individual is selected on the basis of a strategy that doesn't reward the best but punishes the worst. That's the reason why this model has been considered a particular kind of synaptic Darwinism.

Our neuroevolutionary model is also based on a selection strategy. The structural information of our network is not codified by genes. We directly consider the entire network as a population of nodes that can establish connections, generate other units or die. The probability of these events depends on the presence of local and global resources. If there are few resources the population falls, if there is a lot of resources the population grows. Like in the Chialvo and Bak model we don't select the fittest nodes reinforcing their links, but we simply remove the worst nodes when the ecosystem resources are low. This generates a selective process that indirectly rewards the units which can better model the input patterns. Our evolutionary strategy can be seen as a *selective retention* process (Heylighen, 1992) that removes those units which cannot reach a stable state, remaining associated with several input patterns. Even if the stability of a unit is quantified by the minimum distortion error related to it, this information mustn't be considered to be environmental information. The minimum distortion error simply quantifies the difficulty encountered by the unit during the modelling of input patterns.

3. THE EVOLUTIONARY ALGORITHM

Research has confirmed (Roughgarden, 1979; Song and Yu, 1988) that in natural environments the population size along with competition and reproduction rates continuously changes according to some natural resources and the available space in the ecosystem.
These mechanisms have been reproduced in some *evolutionary algorithms*, for example to optimize the evolution of a population of chromosomes in a genetic algorithm (Annunziato and Pizzuti, 2000). We have tried to use a similar strategy for the evolution of a population of units in a self organizing network without using the string representation of genetic programming.
In our model each node is defined by a vector of neighbouring units connected to it, a reference vector and a variable D that is the smallest distance between its reference vector and the closest modelled input. The value of this variable quantifies the *debility* degree of the unit. The lower is D the higher are the chances for the unit to survive. At each presentation of the training input set, D is set to the maximum value. After the presentation of a given input x, if the reference vector w of the unit is modified, the resulting distance between the two vectors ||x-w|| is calculated. If this value is lower than D it becomes its new value.
The training algorithm here used can be subdivided in three phases:

1) Winners are selected. For each input the unit having the closest reference vector is selected.

2) The reference vectors of the winners and their neighbours are updated according to the following formula :

$$(3.1) \quad w(t+1) = w(t) + \alpha(x - w(t)).$$

So the reference vectors w of the selected units are moved towards the relative inputs x of a certain fraction of the distances that separate them. For winners this fraction is two or three orders of magnitude higher than the one used for their neighbours. So winners have the reference vectors moving more quickly towards the inputs.

3) The population of units evolves producing descendants, establishing new connections and eliminating the less performing units. All these events can occur with a well defined probability that depends on the availability of resources.

These rules are iterated until a given goal is achieved. For example the *minimization of the expected quantization error* that is the mean of the distances between the winners and the K inputs they model:

$$(3.2) \quad D = \frac{1}{K} \sum_{i=1}^{K} \|x_i - w_j\|$$

If this value falls below a certain threshold $D_{min}$, training is stopped.

The first two phases can be considered a kind of *winner takes all* strategy, where only the most activated units are selected and enabled to modify their reference vectors. The third phase is the evolutionary phase (fig. 3.1). Each unit i, i=[1…N(t)] where N(t) is the actual population size can meet the closest winner j with probability $P_m$:

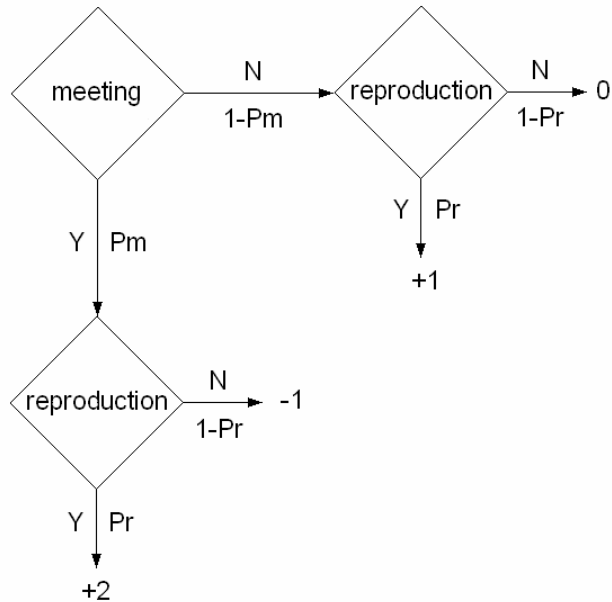

Fig. 3.1 – The evolutionary phase of the algorithm.

If meeting occurs, the two units establish a link and they can interact by reproducing with probability $P_r$. In this case two new units are created. One is closer to the first parent, the other to the second parent:
(3.3)

$$w_1 = \frac{w_{p1} + \frac{w_{p1} + w_{p2}}{2}}{2}$$

$$w_2 = \frac{w_{p2} + \frac{w_{p1} + w_{p2}}{2}}{2}$$

If reproduction doesn't take place due to the lack of resources the weaker unit of the population, i.e. the one with the highest debility degree, is removed.

If unit i doesn't meet any winner it can interact with the closest node k with probability Pr establishing a connection and producing a new unit whose reference vector is set between the parents reference vectors:

$$(3.4) \quad w_2 = \frac{w_{p1} + w_{p2}}{2}$$

When we fix a maximum population size, the ratio between the actual size and the threshold $N(t)/N_{max}$ can be seen as a global resource of the ecosystem affecting the probabilities of the events. For example if the population size is low the reproduction rate should be high. So we can reasonably choose $P_r = 1-N(t)/N_{max}$. If the population size is high, the chance for the units to meet each other will be higher, so we can set $P_m = N(t)/N_{max}$.

We can also consider a local resource that is the ratio between the threshold $D_{min}$ and the debility degree $D_i$ of the unit i. Each unit i should meet a winner with a probability $P_m=(N(t)/N_{max})(1-D_{min}/D_i)$ and $P_r = 1 - P_m$. In this way winners are not encouraged to migrate to other groups of nodes and weaker units don't participate in reproduction activities.

We can estimate the population grow rate in the following way:
(3.5)

$$N(t+1) = N(t) + 2P_m P_r N(t) - P_m(1-P_r)N(t) + (1-P_m)N(t) - (1-P_m)P_d N(t) =$$
$$= 2N(t) - 2P_m^2 N(t) =$$
$$= 2N(t)\left(1 - \frac{N(t)^2}{M_P^2}\right) \Rightarrow X(t+1) = 2X(t)\left(1 - X(t)^2\right) \quad (first \quad model)$$
$$= 2N(t)\left(1 - \frac{N(t)^2}{M_P^2}\left(1 - \frac{D_{min}}{D}\right)^2\right) \Rightarrow X(t+1) = 2X(t)\left(1 - X(t)^2\left(1 - \frac{D_{min}}{D}\right)^2\right) \quad (second \quad model)$$

where X(t) is the normalized size $N(t)/N_{max}$. This is the *quadratic-logistic* map of Annunziato and Pizzuti(Annunziato and Pizzuti, 2000):

$$(3.6) \quad X(t+1) = aX(t)\left(1 - X(t)^2\right)$$

They proved that by varying the parameter different chaotic regimes arise. For a<1.7 the behaviour is not chaotic, for 1.7<a<2.1 we have chaotic regimes with simple attractors localized in a fixed part of the plane of the phases. Theoretically for the first model we expect to obtain a chaotic regime that is described by a simple attractor. In the second model the factor $(1 - D_{min}/D)$ might reduce the influence of the negative feedback in the final part of network training.

It is possible to demonstrate that during the evolution the population size converges to $N(t) = 0.72 N_{max}$. In this phase the probability that a unit establishes n connections with the other ones for the first model, considering only clusters of n units, is given by:

$$(3.7) \quad P(n) = \left(\frac{0.72 N_{max}}{N_{max}}\right)^n - \sum_{i=1}^{0.72 N_{max}-1-n} \left(\frac{0.72 N_{max}}{N_{max}}\right)^{i+n} = \alpha n^{-\beta}$$

It has to be pointed out we have subtracted the probability that such n links developed within a cluster of more than a n unit.

The coefficients $\alpha$ and $\beta$ of the power law are considered constant at the end of the training. To compute their values, we can take into consideration the cases n=1 and n=0.72N $N_{max}$-1 which correspond to the minimum and maximum number of connection at the end of the training.

$$(3.8) \quad P(1) = 0.72 - \sum_{i=1}^{0.72N_{max}-2} 0.72^{i+1} = \alpha 1^{-\beta} = \alpha$$

$$P(0.72N_{max} - 1) = 0.72^{0.72N_{max}-1} = \left(0.72 - \sum_{i=1}^{0.72N_{max}-2} 0.72^{i+1}\right)(0.72N_{max} - 1)^{-\beta}$$

$$\Rightarrow \beta = \log_{0.72N_{max}-1}\left(\frac{0.72 - \sum_{i=1}^{0.72N_{max}-2} 0.72^{i+1}}{0.72^{0.72N_{max}-1}}\right)$$

The distribution tail of the degrees tends to stretch when the maximum size of the population increases, it means that in wider networks there are more hubs with a higher degree.
For the second model we can consider that at the end of the training (1-Dmin/D) ~ $\varepsilon$
So the probability that a unit establishes n links becomes:
(3.9)

$$P(n) = \left(\frac{0.72N_{max}}{N_{max}}\varepsilon\right)^n - \sum_{i=1}^{0.72N_{max}-1-n}\left(\frac{0.72N_{max}}{N_{max}}\varepsilon\right)^{i+n} = \alpha n^{-\beta}$$

$$\Rightarrow \beta = \log_{0.72N_{max}-1}\left(\frac{0.72\varepsilon - \sum_{i=1}^{0.72N_{max}-2}(0.72\varepsilon)^{i+1}}{(0.72\varepsilon)^{0.72N_{max}-1}}\right),$$

and the considerations made for the first model can be therefore extended to the second model.

4. TRAINING THE NET: SIMULATIONS

We have compared the performances of our networks with those of a Growing Neural Gas in categorizing bidimensional inputs.
GNG is a self organizing network which thanks to both the competitive hebbian learning strategy and the neural gas algorithm can categorize inputs without altering their exact dimensionality. For the GNG, the parameters of the model are $\alpha = 0.5$, $\beta = 0.0005$ and at each $\lambda = 300$ steps a new unit is inserted. The maximum age of the links is set to 88.
For the two different ENG models, the parameters are $\alpha = 0.05$, $\beta = 0.0006$ and the maximum size is set to $N_{max} = 120$.

As stopping criterion for both the algorithms we have chosen the *minimization of the expected quantization error* that is the average distance between the winners and the corresponding inputs.

We have considered two different input domains. In the first case inputs are localized within four square regions, in the second one inputs are uniformly distributed in a ring region.

As shown in fig.4.1 after the training, GNG reference vectors are all positioned in the input domain. In the Evolutionary Self Organizing Networks (fig.4.2a and fig.4.2b) some units fall outside the input domain, but in this way the network remains fully connected. The nodes' distribution statistical analysis confirms what appears to be intuitively patent: the emerging network structure is a typical scale-free one, i.e. a structure where few hubs manage the links.

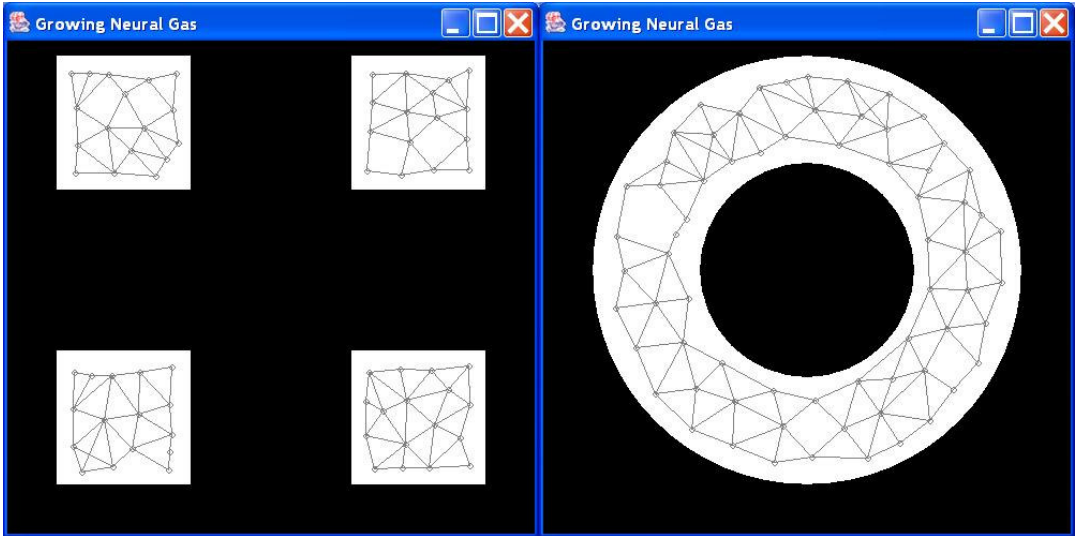

Fig. 4.1 – Growing Neural Gas simulations.

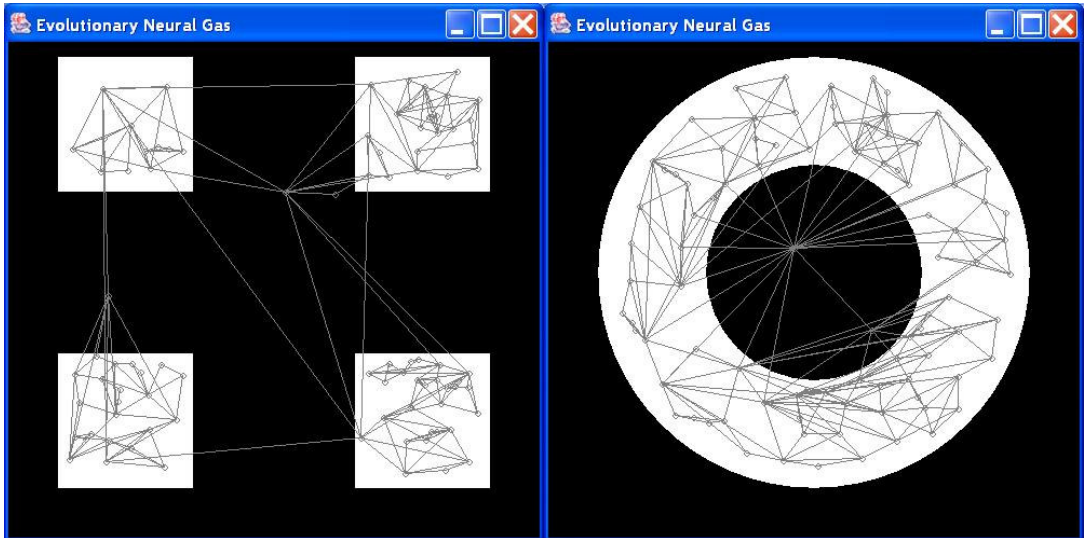

Fig. 4.2 a – Evolutionary Self Organizing Network simulations (first model).

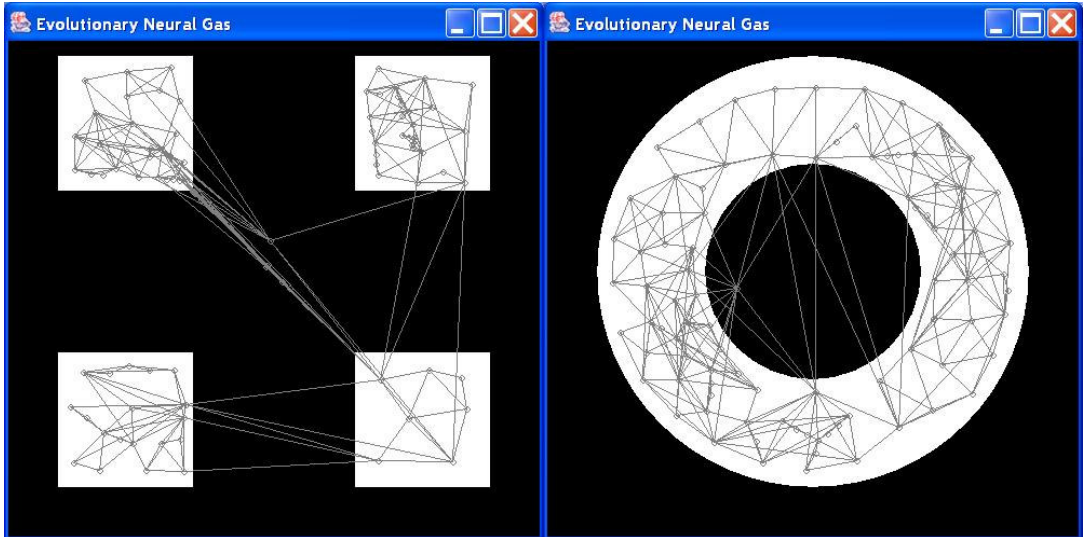

Fig. 4.2 b – Evolutionary Self Organizing Network simulations (second model).

We trained 30 networks of each type obtaining the average degree distributions reported in fig.4.3-4.5. In tab. 4.1 – 4.2 are reported the average values of the structural parameters of the two networks.

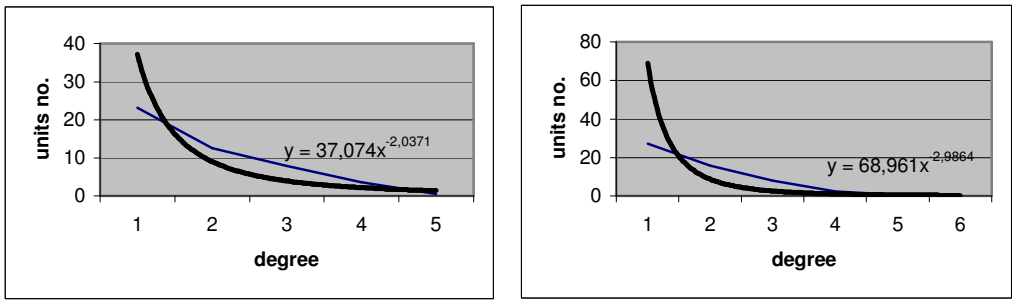

Fig. 4.3 – Average degree distribution in GNG (two different input domains)

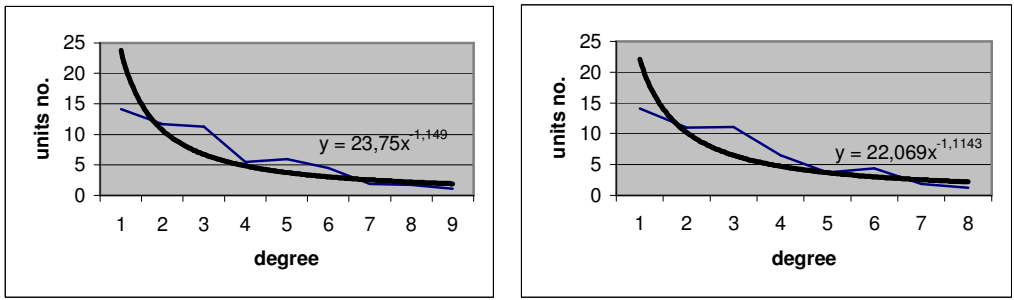

Fig. 4.4 – Average degree distribution in ENG (first model, two different input domains)

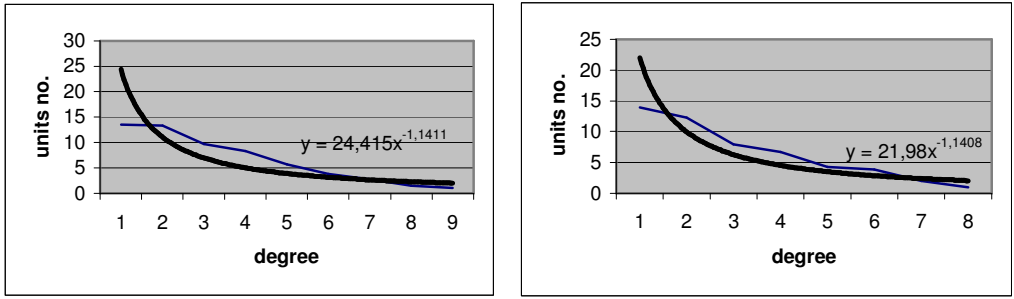

Fig. 4.5 – Average degree distribution in ENG (second model, two different input domains)

While GNG have a high value for the average path length and a low clustering coefficient, ENG have a short average path length and a high clustering coefficient which along with the power law tail of the degree distribution confirm its scale free graph features.

|  | Average path length | Clustering coefficient | Power law exponent |
|---|---|---|---|
| GNG | - | 0.49 | 2.04 |
| ESON (1$^{st}$) | 3.82 | 0.64 | 1.15 |
| ESON(2$^{nd}$) | 3.92 | 0.63 | 1.14 |

Tab. 4.1 – Comparison of structural parameters (average values, first input domain)

|  | Average path length | Clustering coefficient | Power law exponent |
|---|---|---|---|
| GNG | 6.4 | 0.42 | 2.98 |
| ESON (1$^{st}$) | 3.61 | 0.58 | 1.11 |
| ESON(2$^{nd}$) | 3.67 | 0.59 | 1.14 |

Tab. 4.2 – Comparison of structural parameters (average values, second input domain)

Fig. 4.6 – 4.7 shows the population dynamics of the two ENG models.
The structure shared by the two different ENG models is due to the fact that the winner units tend to establish the greatest number of connections. These are the favoured units with which each node try to establish a connection. If the probability depends also on the local distortion error as it happens in the second model, we obtain a final structure that is more similar to the GNG, which is to say more similar to a gas. In point of fact, the conditions to create a new link become more restrictive, reducing the interaction among each cluster and the whole network. The structure of connections seems to extend more uniformly in the regions where inputs are present as it can be seen in picture 4.2b (more evident in the circular distribution).
Picture 4.7 shows the dynamics of the populations in the two different models of ENG. In the first model the population size seems to converge to the final value of $0.72N_{max}$, confirming the experimental results of Annunziato and Pizzuti. As it can be noticed in fig. 4.6, since the d value gradually diminishes during the training, the influence of the factor *(1-$D_{min}$/d)* grows reducing the effects of the negative feedback which characterizes the quadratic logistic map. This justifies the sudden growth of the population at the end of the training in the second model.

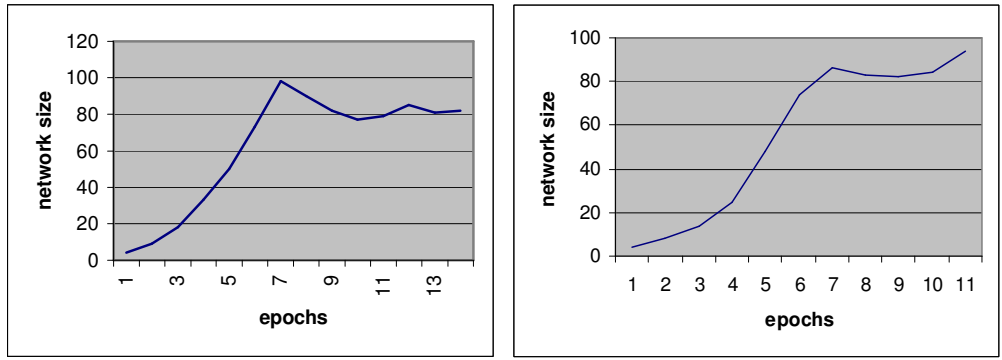

Fig. 4.6 – Network size evolution of the two ENG models (first input domains).

At the end of training new units connect with the winner units which have a lower d, while the subgroups of units become more isolated. Considering the function (X(t),X(t+1)) the attractor becomes more marked in the second model. This means that the system tends to converge more toward a precise final state with a lower interaction among the groups of units.

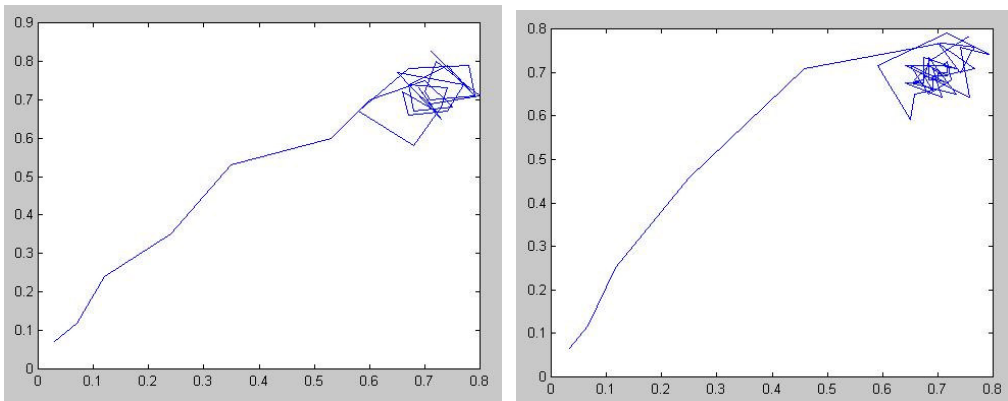

Fig. 4.7 – population dynamics (X(t),X(t+1)) of the two Esonet models (first input domain).

## 5. THE ROLE OF INFORMATION IN FUNCTIONAL SPECIALIZATION AND INTEGRATION

We can classify a system as complex when it is made up of different parts heterogeneously interacting. In addition, its behaviour and its structure have to be neither completely casual (as it happens in a gas) nor too regular (as it happens in a crystal). In Nature we generally observe the co-existence of functionally highly specialized integrated areas.

That's what happens in the brain, where different areas and groups of neurons interact to give rise to an integrated and unitary cognitive scenario (G. M. Edelman, G. Tononi, 2000).

Edelman has introduced the integration, reciprocal information and complexity concepts in order to mathematically define the functional organization of the cerebral structures.

Within a complex system, a subset of elements can be defined an integrated process if – on a given temporal scale – the elements interact more strongly with each other than with the system. In a neural net or in a self-organizing one it means that the units of an integrated group will tend to simultaneously activate themselves.

When the units in a subset are independent, the system's entropy reaches its maximum value which is the sum of the entropies of the single elements (local entropies). On the contrary, when any kind of interaction occurs, the global entropy decreases so becoming lower than the sum of the local entropies. The integration measure is, therefore, a natural indicator of the system informational "capacity".

So the integration of a subset of network units can be calculated by deducting the sum of the entropies of each single component $(x_i)$ from the entropy of the system considered as a whole. If each unit can only take two states (activated/not-activated), the amount of the possible activation patterns of a subset with N units is $2^N$. So the system maximum entropy is:

$$(5.1) \quad H_{max}(X) = \sum_{i=1}^{n} H(x_i) = \sum_{i=1}^{n} p_i \log_2\left(\frac{1}{p_i}\right) = \frac{2^N}{2^N} \log_2\left(1 \middle/ \frac{1}{2^N}\right) = \log_2 2^N = N$$

and the integration will be:

$$(5.2) \quad I(X) = \sum_{i=1}^{n} H(x_i) - H(X)$$

for the self-organized net here considered, the integration of a sub-group of units takes the following expression:

$$(5.3) \quad I(X) = N - \sum_{i=1}^{N-1} \binom{N}{i+1} P_i \log_2\left(\frac{1}{P_i}\right)$$

where $P_i$ is the probability for a node to establish $i$ connections. The overall number of the system' states is equal to the total number of possible groups of $i+1$ units. Groups of units having the same dimension (groups of i+1 units) give the same contribution to the entropy of the system.

If we choose the WTA strategy as activation modality, for each presented input only a single unit (the winner) and the 1<i<N-1 $i$ units will activate themselves. All the other ones remain not-activated.

The probability for a node to create connections is ruled by the power law $P_i = \alpha k^{-\beta}$, with $\alpha$ and $\beta$ depending on 1) the network dimension, 2) the local distortion errors (for the second model) and 3) the particular evolution of the network structure, i.e. the dynamic behaviour of $\alpha(t)$ and $\beta(t)$.

So the integration of the two self-organizing network here presented is:

$$(5.4) \quad I(X) = N - \sum_{i=1}^{N-1} \binom{N}{i+1} (\alpha i^{-\beta})^i \log_2\left(\frac{1}{\alpha i^{-\beta}}\right)^i$$

The integration can be seen as a measure of the statistic dependency within a subset of units. The stronger their interactions are, the higher their integration.

In order to measure the statistic dependency between a subset and the whole system, Edelman introduced the concept of mutual information. Given an $n$ subset made up of $k$ elements $(X_n^k)$ and its complement in the system $(X - X_n^k)$, the mutual information is:

$$(5.5) \quad IR(X_n^k; X - X_n^k) = H(X_n^k) + H(X - X_n^k) - H(X)$$

The mutual information is essential to evaluate the differentiation degree of a system, i.e. it is a significant index of the system' "resolution" degree, calculated on the subdividable and distinct states.

In order to measure the information of an integrated activation pattern, we calculate how the states of a given subset can differentiate them from the whole system ones. Which thing, following Edelman, is equivalent to considering the whole system as the observer of itself. In fact, if entropy measures the variability of a system according to an external observer evaluation, the mutual information measures the system variability according to an observer ideally placed within the system itself.

The overall measure of the differentiation degree of a complex system is given by the mutual information average between each subset and the whole system:

$$(5.6) \quad C(X) = \sum_{k=1}^{N/2} \langle IR(X_n^k; X - X_n^k) \rangle$$

Edelman defined such measure as complexity and its value is high if each subset can averagely take many different states which are statistically depending on the whole system's ones, so it shows how the system is differentiated. High complexity values correspond to an optimal synthesis of functional specialization and functional integration. Systems whose elements are not integrated (such as a gas) or not specialized (such as an homogeneous crystal) have a minimum complexity.

In the evolutionary neural gas case, the WTA strategy limits the integration among the activation patterns. So the mutual information between any activation pattern and the other possible patterns is equal to zero. It justifies the use of the term "gas", since the patterns behave like isles of information weakly interacting each other.

If there were selected more winner units for the same input signal in the early training phase, we could get a given system status characterized by $i + 1$ activated units not only by the activation of just a single winner and its related $i$ units, but also by the activation of more winners. therefore we should also take into consideration all the possible subgroups with $j+1$ elements.

The mutual information formula between a subgroup with $k$ activated units and the system is given by:

(5.7)

$$H(X_n^k) = \sum_{i=1}^{k-1} \binom{k}{i+1} \left[ (\alpha i^{-\beta})^i + \sum_{j=1}^{i} \binom{i+1}{j+1} (\alpha j^{-\beta})^j \right] \log_2 \left( \frac{1}{(\alpha i^{-\beta})^i + \sum_{j=1}^{i} \binom{i+1}{j+1} (\alpha j^{-\beta})^j} \right)$$

$$H(X - X_n^k) = \sum_{i=1}^{k-2} \binom{N}{i+1} \left[ (\alpha i^{-\beta})^i + \sum_{j=1}^{i} \binom{i+1}{j+1} (\alpha j^{-\beta})^j \right] \log_2 \left( \frac{1}{(\alpha i^{-\beta})^i + \sum_{j=1}^{i} \binom{i+1}{j+1} (\alpha j^{-\beta})^j} \right) +$$

$$+ \left( \binom{N}{k} - 1 \right) \left[ (\alpha (k-1)^{-\beta})^{k-1} + \sum_{j=1}^{k-1} \binom{k}{j+1} (\alpha j^{-\beta})^j \right] \log_2 \left( \frac{1}{(\alpha (k-1)^{-\beta})^{k-1} + \sum_{j=1}^{k-1} \binom{k}{j+1} (\alpha j^{-\beta})^j} \right) +$$

$$+ \sum_{i=k}^{N-1} \binom{N}{i+1} \left[ (\alpha i^{-\beta})^i + \sum_{\substack{j=1 \\ j \neq k-1}}^{i} \binom{i+1}{j+1} (\alpha j^{-\beta})^j + \left( \binom{i+1}{k} - 1 \right) (\alpha j^{-\beta})^j \right].$$

$$\log_2\left(\frac{1}{\left(\alpha i^{-\beta}\right)^i + \sum_{\substack{j=1 \\ j \neq k-1}}^{i} \binom{i+1}{j+1}\left(\alpha j^{-\beta}\right)^j + \left(\binom{i+1}{k}-1\right)\left(\alpha j^{-\beta}\right)^j}\right)$$

$$H(X) = \sum_{i=1}^{N-1} \binom{N}{i+1}\left[\left(\alpha i^{-\beta}\right)^i + \sum_{j=1}^{i}\binom{i+1}{j+1}\left(\alpha j^{-\beta}\right)^j\right]\log_2\left(\frac{1}{\left(\alpha i^{-\beta}\right)^i + \sum_{j=1}^{i}\binom{i+1}{j+1}\left(\alpha j^{-\beta}\right)^j}\right)$$

To provide the system with a greater level of complexity, in order to favouring the integration among the network unit subgroups, it is, therefore, necessary adopting a strategy different from the WTA in the early training phases so as to select more winner units.

### 6) CONCLUSIONS AND FUTURE WORKS

The here presented self-organizing network can be considered as an example of autopoietic system which evolves by means of a closed network of interactions and based upon the production of components (the categorization units). In the course of the reproductive dynamics, those ones produce other components, also belonging to the system (i.e. other categorization units) which maintain the system identity over time with respect to the experimental task.

In particular, it has to be noticed that they are not just the environmental information to lead the evolution of the network of connections, but rather the network internal status, which is individuated globally by the size that the population has reached and locally by the values of the parameters of the units. The latter show the difficulties that the units encounter in modelling the presented input, such difficulty is directly proportional to the amount of variations their reference vectors are subjected to.

Learning and the capability to model the system external inputs, therefore, emerges more by means of the population internal dynamics than by means of a learning algorithm.

The appearing of a scale-free structure emerging from the choice of the population dynamics is peculiarly significant for the model's biological plausibility. Which thing describes a quite phase-transition-like status where cluster "float" as informational "isles" in a "gaseous" configuration. It is worthy noticing that the WTA strategy and the environmental noise (probabilistic laws) suffice to create a kind of basic informational skeleton around which more interconnected functional structures can then aggregate. In the nervous system, it plausibly happens according to an essentially genetic design. Such kind of neural dynamics guarantees flexibility and redundancy to the informational nuclei which are ready to synchronize and connect through signals. Actually, what we tried here to describe is a proto-neural scenario with low integration of clusters which are specialized in easy categorization tasks.

Developing the ENG model requires to investigate different synchronization scenarios among clusters and their ensuing functional integration to execute more complex tasks. In particular, it is necessary to modify the evolutive dynamics so as to mane the connections among units active. In this way, it should be possible to create a dynamic neural topology susceptible of hierarchical organization.

Everything seems to confirm not only the deep reasons for the scale-free structures recurring in nature (Z. Toroczkai, K. E. Bassler, 2004), but also the fundamental lesson associating complexity with a thin border zone between integration and differentiation among the functional modules of a system.

Acknowledgements: The authors thank Eliano Pessa and Graziano Terenzi for their precious suggestions.